\documentclass[%
 reprint,
 amsmath,amssymb,
 aps,
]{revtex4-2}
\usepackage{amsmath}
\usepackage{graphicx}
\usepackage{dcolumn}
\usepackage{physics}
\usepackage{bm}
\usepackage{amssymb}
\usepackage[verbose]{placeins} 
\usepackage{xcolor}
\begin{document}

\preprint{APS/123-QED}

\title{Quantum Decay of an Optical Soliton}

\author{Stuart Ward$^{1,2}$}
\author{Rouzbeh Allahverdi$^1$}%
\author{Arash Mafi$^{1,2}$}
\email{mafi@unm.edu}
\affiliation{$^1$Department of Physics \& Astronomy, University of New Mexico, Albuquerque, NM 87106, USA.\\
$^2$Center for High Technology Materials, University of New Mexico, Albuquerque, NM 87106, USA.
}
\date{\today}

\begin{abstract}
Optical solitons are known to be classically stable objects which are robust to perturbations. In this work, we show that due to quantum mechanical effects, an optical soliton that is initially in a classical soliton coherent state will shed photons into the continuum and hence decay. The standard formulation of the quantized soliton uses the linearized version of the quantum nonlinear Schr\"{o}dinger equation in the background of the classical soliton, and the quantized soliton remains stable in this approximation. We show that if higher-order interaction terms are taken into account, the soliton is no longer stable, and its photon number decreases quadratically as a function of the number of soliton cycles. We compute the power spectrum for the continuum radiation and find a narrow band that is localized about the initial soliton momentum with a cut-off that is inversely proportional to the initial soliton width.
\end{abstract}

\maketitle

\section{Introduction}
When light with sufficient intensity propagates within a dielectric waveguide, the refractive index is modified due to the Kerr effect, resulting in a nonlinear equation for the propagating optical field~\cite{HH,Boyd,Agarwal,Powers}. In the slowly varying envelope approximation, the resulting wave equation is known as the Nonlinear Schr\"{o}dinger Equation (NLSE), which gives rise to localized solutions known as {\em solitons}~\cite{Alan,Drazin}. Optical solitons have many applications in photonics; for example, the field of telecommunications has utilized the dispersionless nature of solitons in long distance data transmission~\cite{Hasegawa,Nakazawa}, and the field of integrated photonics has recently leveraged soliton microcombs in various applications~\cite{Shen,Guidry1,chembo}. Furthermore, optical systems have been used as models for black hole analogues, with optical solitons having been used to model Hawking radiation ~\cite{Raul,Linder} -- a phenomenon which is known more generally as quantum evaporation. Optical quantum soliton evaporation has previously been studied by way of computing an approximated power spectrum for a soliton which is initially in a fundamental soliton state~\cite{Conti}. Similarly, geometric approaches have been utilized in calculating the temperature of an optical soliton ~\cite{Robson,Villari}. We would also like to highlight the extensive work of Malomed and collaborators regarding the analysis of decaying optical and non-optical solitons subject to perturbations in classical and quasiclassical quantized frameworks~\cite{Malomed1,Malomed2,Malomed3,Malomed4,Malomed5,Malomed6,Malomed7,Malomed8}.

Here, we study the evaporation of a quantum mechanical soliton using the quantum NLSE in the background of the classical soliton. Our approach is analogous to the techniques used in the standard model of particle physics, where the Higgs field is expanded around a constant classical expectation value, except that the classical soliton is both time- and space-dependent~\cite{Higgs}. At the linear level, the resulting theory in the background of the soliton has exact solutions in the form of four bound-states, as well as continuum states~\cite{Haus1,Lai1,Haus2,Haus3,HausYu}. These bound-states characterize the modified soliton parameters. The bound-state and continuum solutions evolve trivially in time in the linearized theory. To observe more interesting dynamics, we include the higher-order interaction terms that couple the bound-states to the continuum. The higher-order contribution acts as a perturbation on the linearized theory and prompts the soliton to lose photons to the continuum, resulting in soliton evaporation. The rigorous treatment of this process is the subject of this paper.

The rest of this paper is organized as follows. In Section II, we review the linearized theory of the quantum soliton. In Section III, we perform a perturbative approach on the linearized theory to show that the soliton's photon number does indeed decrease due to quantum effects. In Section IV, we compute the power spectrum of the generated continuum radiation as a function of the number of soliton cycles, and calculate the band-width of the radiation spectrum. In the Appendix, a more comprehensive review of the linearized NLSE is given, along with a list of relevant vacuum state expectation values and other supporting mathematical expressions.

\section{Linearized Theory of the Quantum Soliton}
In this section, an overview of the essential parts of the linearized field approximation applied to the quantized NLSE is given~\cite{Haus1,Lai1,Kaup1,Gordon1} -- a more complete review is provided in Appendix A. The linearized theory of the quantized NLSE allows one to treat the perturbations of the soliton parameters, along with the generated continuum radiation, as quantum operators. More precisely, the soliton perturbation operators are given as the change in the four soliton parameters, namely the photon number ($\Delta \hat{n}_{0}$), phase ($\Delta \hat{\theta}_{0}$), position ($\Delta \hat{x}_{0}$), and momentum ($\Delta \hat{p}_{0}$). The accompanied continuum radiation is denoted as $\Delta \hat{v}_{c}$.

The equation of motion describing the propagation of a quantum soliton field envelope in a single-mode optical waveguide (in the co-moving frame of the soliton) is given by the following equation, known as the quantized Nonlinear Schr\"{o}dinger equation~\cite{Huttner,Haus4,Haus5,Wright,Yao,Kartner2,Dvali}:
\begin{equation}
    \label{eq:QNLSE}
    i\frac{\partial}{\partial t}\hat{\phi}(x,t) = -\frac{\partial^2}{\partial x^2}\hat{\phi}(x,t) - 2|c|\hat{\phi}^{\dagger}(x,t)\hat{\phi}(x,t)\hat{\phi}(x,t).
\end{equation}
The quantum field operator, $\hat{\phi}(x,t),$ obeys the usual commutation relations~\cite{Matsko}:
\begin{subequations}
\label{eq2}
\begin{align}
&[\hat{\phi}(x,t),\hat{\phi}^{\dagger}(y,t)] = \delta(x-y),\\
&[\hat{\phi}(x,t),\hat{\phi}(y,t)] = [\hat{\phi}^{\dagger}(x,t),\hat{\phi}^{\dagger}(y,t)] = 0.
\end{align}
\end{subequations}
The number operator is given by
\begin{equation}\label{NumberOperator}
    \hat{N} = \int dx\hat{\phi}^{\dagger}(x,t)\hat{\phi}(x,t),
\end{equation}
and is a conserved quantity of the above theory.

We are interested in the evolution of the quantum fluctuations about the classical soliton background. These fluctuations are viewed as perturbative in the sense that the even-moment expectation values of the fluctuation operator are small in comparison to the even-moments of the classical soliton solution. The quantum field operator may be redefined as the expansion about the classical soliton solution as follows:
\begin{equation}\label{Perturbed}
    \hat{\phi}(x,t) = \phi_{cl}(x,t) + \hat{v}(x,t)\exp[{{i\frac{n_{0}^{2}|c|^{2}}{4}t}}],
\end{equation}
where
\begin{equation}\label{classicalsol}
\begin{split}
    &\phi_{cl}(x,t) =\frac{n_{0}|c|^{1/2}}{2}\sech\Big(\frac{n_{0}|c|}{2}(x-x_{0}-2p_{0}t)\Big)
    \\
   & \times \exp\Big[i\frac{n_{0}^{2}|c|^{2}}{4}t - ip_{0}^{2}t + ip_{0}(x-x_{0}) + i\theta_{0}\Big].
    \end{split}
\end{equation}
The  constants $n_{0}$, $p_{0}$, $x_{0}$, and $\theta_{0}$ represent the initial photon number, momentum, position, and phase. We take  $p_{0} = x_{0} = \theta_{0} = 0$ for the remainder of this work. Note that the vacuum state for the quantum fluctuation field operator, $\hat{v}(x,t)$, in Eq.~(\ref{Perturbed}) at $t=0$ is the classical soliton coherent state.

From Eq.~(\ref{eq2}) and Eq.~(\ref{Perturbed}), the quantum fluctuation operator commutation relations take the form
\begin{subequations}
\label{commutation2}
\begin{align}
&[\hat{v}(x,t),\hat{v}^{\dagger}(y,t)] = \delta(x-y),\\
&[\hat{v}(x,t),\hat{v}(y,t)] = [\hat{v}^{\dagger}(x,t),\hat{v}^{\dagger}(y,t)] = 0.
\end{align}
\end{subequations}
Inserting Eq.~(\ref{Perturbed}) into Eq.~(\ref{eq:QNLSE}) and keeping terms up to first order in $\hat{v}(x,t)$, we arrive at the following equation of motion:
\begin{equation}\label{eq:linNSLE_relabeled}
\begin{split}
    &\frac{4}{n^{2}_{0}|c|^{2}}\frac{\partial}{\partial t}\hat{v}(x',t) = i\Big(\Big[\frac{\partial^2}{\partial x'^2} - 1\Big]\hat{v}(x',t)\\ &+2\sech^{2}(x')\big[2\hat{v}(x',t)+\hat{v}^{\dagger}(x',t)\big]\Big),
    \end{split}
\end{equation}
where we have used the following rescaling for convenience
\begin{equation}\label{eq:relabel}
\begin{split}
&x' = \frac{n_{0}|c|x}{2}.
\end{split}
\end{equation}
Unless explicitly mentioned, from here on we drop the prime notation in $x^\prime$; the proper scaling factors will be taken into account when necessary.

We utilize a doublet notation~\cite{Kartner1} for the quantum fluctuation operator and its adjoint, which allows Eq.~(\ref{eq:linNSLE_relabeled}) to be written in the following compact form:
\begin{equation}\label{eigen}
    \frac{4}{n^{2}_{0}|c|^{2}}\frac{\partial}{\partial t}\Vec{v} = \hat{L}\Vec{v},
\end{equation}
such that
\begin{equation}
    \Vec{v} = \begin{bmatrix} \hat{v}(x,t) \\ \hat{v}^{\dagger}(x,t) \end{bmatrix}
\end{equation}
and
\begin{equation}\label{linearop}
   \hat{L} = i\sigma_{3}\big[(\frac{\partial^2}{\partial x^2} - 1)+2\sech^{2}(x)(2+\sigma_{1})\big],
\end{equation}
where $\sigma_{1}$ and $\sigma_{3}$ are the usual Pauli matrices.

The fluctuation operator may be expanded in the following form:
\begin{equation}\label{expand}
\begin{split}
    &\Vec{v}(x,t) = \\&\frac{\partial \Vec{\phi}_{cl}(x,t)}{\partial n_{0}}\rvert_{t=0}\Delta \hat{n}_{0}(t)+\frac{\partial \Vec{\phi}_{cl}(x,t)}{\partial \theta_{0}}\rvert_{t=0}\Delta \hat{\theta}_{0}(t)\\
    &+\frac{\partial \Vec{\phi}_{cl}(x,t)}{\partial p_{0}}\rvert_{t=0}\Delta \hat{p}_{0}(t) +\frac{\partial \Vec{\phi}_{cl}(x,t)}{\partial x_{0}}\rvert_{t=0}\Delta \hat{x}_{0}(t) \\ &+ \Delta \Vec{\hat{v}}_{c}(x,t).
    \end{split}
\end{equation}
The first four operators of the above expression represent the perturbed parameters of the original soliton solution, while the last term represents the continuum radiation field. The physical motivation for the above expansion lies in the observation that the classical soliton is a stable object and a perturbation results in the soliton returning to a soliton configuration (now with modified photon number, phase, momentum, and position) along with the generation of radiation which is shed into the continuum.

With the field redefinition given in Eq.~(\ref{Perturbed}), the number operator given in Eq.~(\ref{NumberOperator}) now takes the form:
\begin{equation}\label{linnumb}
    \hat{N} = n_{0}+\Delta\hat{n}_{0}(t)+\frac{2}{n_{0}|c|}\int dx\hat{v}^{\dagger}(x,t)\hat{v}(x,t),
\end{equation}
where
\begin{equation}
       \Delta \hat{n}_{0}(t) = \frac{1}{|c|^{1/2}}\int_{-\infty}^{\infty}dx \sech{(x)}\big(\hat{v}(x,t)+\hat{v}^{\dagger}(x,t)\big),
\end{equation}
and
\begin{eqnarray}\label{extra}
    \hat{v}^{\dagger}(x,t)\hat{v}(x,t) & = & \Delta\hat{v}^{\dagger}_{sol}\Delta\hat{v}_{sol}+\Delta\hat{v}^{\dagger}_{sol}\Delta\hat{v}_{c} \, \nonumber \\ & + & \Delta\hat{v}^{\dagger}_{c}\Delta\hat{v}_{sol}+\Delta\hat{v}^{\dagger}_{c}\Delta\hat{v}_{c} \, , 
\end{eqnarray}
with
\begin{equation}\label{vsol}
\begin{split}
    \Delta\hat{v}_{sol}(x,t) =& \phi_{n_{0}}(x,0)\Delta \hat{n}_{0}(t)+\phi_{\theta_{0}}(x,0)\Delta \hat{\theta}_{0}(t)+\\&\phi_{p_{0}}(x,0)\Delta \hat{p}_{0}(t)+\phi_{x_{0}}(x,0)\Delta \hat{x}_{0}(t),
    \end{split}
\end{equation}
and $\Delta\hat{v}_{c}$ defined below in Eq.~(\ref{lintimecon}).

Note that Eq.~(\ref{linnumb}) is separated into the initial soliton photon number ($n_{0}$), the change in soliton photon number ($\Delta \hat{n}_{0}(t)$), and four different couplings involving the change in soliton parameters ($\Delta\hat{v}_{sol}$) and the continuum radiation ($\Delta\hat{v}_{c}$). 

As can be seen from Eq.~(\ref{evolpara}) of Appendix A, at the linear order the change in photon number operator evolves independently of time and is given as 
\begin{equation}
\Delta \hat{n}_{0}(t) = \Delta\hat{n}_{0}(0).
\end{equation}
Similarly, Eq.~(\ref{evolpara}) shows that the continuum operator evolves as 
\begin{align}\label{lintimecon}
        &\Delta \hat{v}_{c}(x,t) = \\ 
        \nonumber
        &\int_{-\infty}^{\infty}dk\, e^{-ikx}\big[(k-i\tanh(x))^{2}\hat{a}(k,t)+\sech^{2}(x)\hat{b}(k,t)],
\end{align}
where $\hat{a}(k,t)$ and $\hat{b}(k,t)$ are given in Eq.~(\ref{contimeevolve}).

We see that at the linear order, the continuum evolves as a free-field - that is to say, with all of the time dependence isolated in the phase; as can be seen in Eq.~(\ref{contimeevolve}). Furthermore, we see that the continuum portion of the field and the four soliton parameters evolve independently of one another. In order to see mixing between the continuum and soliton parameters and hence to study more complicated dynamics, we must examine the higher order terms which were excluded during the linearization process. In the next section, we do this by taking the second order corrections into account.

\section{Soliton Decay}
In this section, we use an iterative perturbation approach to calculate the expectation value and variance for the change in photon number of the soliton at second order in $\hat{v}$~\cite{Kaup2}. The change in photon number is given as a function of the number of soliton cycles. We show that the change in photon number is negative and hence the soliton does indeed decay.

We now take into account the second order contributions that were neglected during the derivation of  Eq.~(\ref{eq:linNSLE_relabeled}). Again, inserting Eq.~(\ref{Perturbed}) into Eq.~(\ref{eq:QNLSE}), but now keeping terms up to second order in $\hat{v}$, and continuing with the doublet notation, results in the perturbation operator evolving as:
\begin{equation}\label{NL}
    \frac{4}{n^{2}_{0}|c|^{2}}\frac{d}{dt}\Vec{v} = \hat{L}\Vec{v} + \widehat{NL}\Vec{v},
\end{equation}
where
\begin{equation}
    \widehat{NL} = \frac{4i}{n_{0}|c|^{1/2}}\sech(x)\begin{bmatrix}
2\hat{v}^{\dagger}+\hat{v} & 0 \\
-2\hat{v}^{\dagger} & -\hat{v}^{\dagger}
\end{bmatrix}.
\end{equation}
Using the orthogonality relations provided in Appendix A, the time evolution for the change in photon number operator is now given as:
\begin{equation}\label{photonnumber}
    \frac{d}{dt}\Delta \hat{n}_{0}(t) = in_{0}|c|\int_{-\infty}^{\infty}dx\,\sech^{2}{(x)}\big(\hat{v}^{2}-\hat{v}^{\dagger\,2}\big).
\end{equation}
We solve this equation by way of iterative perturbation. Using Eq.~(\ref{lintime}) and Eq.~(\ref{lintimecon}), the first order linear solution ($\hat{v}_{lin}(x,t)$) is given as:
\begin{equation}\label{timeevol}
\begin{split}
   &\hat{v}_{lin}(x,t) = \big[ \big (\phi_{n_{0}}\Delta \hat{n}_{0}(0)+\phi_{\theta_{0}}\Delta \hat{\theta}_{0}(0)\\ &+\phi_{p_{0}}\Delta \hat{p}_{0}(0)+ \phi_{x_{0}}\Delta \hat{x}_{0}(0)\big) + 
    \Delta \hat{v}_{c}(x,t) \\ &+ \big(\frac{n_{0}|c|^{2}}{2}\phi_{\theta_{0}}\Delta \hat{n}_{0}(0) + 2\phi_{x_{0}}\Delta \hat{p}_{0}(0)\big)t\big].
    \end{split}
\end{equation}
Inserting Eq.~(\ref{timeevol}) into the right hand side of Eq.~(\ref{photonnumber}) and taking the expectation value of both sides results in:
\begin{equation}\label{timeexp}
\begin{split}
    &\frac{d}{dt}\big<\Delta \hat{n}_{0}(t)\big>  = \\&-2n_{0}\,|c|\,\text{Im}\Big[\int_{-\infty}^{\infty}dx\, \sech^{2}{(x)}\big<\hat{v}^{2}_{lin}(x,t)\big>\Big].
    \end{split}
\end{equation}

Using the expectation values given in the Appendix B, along with evaluating at $t_{m} = (8\pi m)/(n_{0}^{2}|c|^{2})$, one arrives at the following quantity for the change in photon number of the soliton after m-cycles of the soliton phase:
\begin{equation}\label{Solution}
    \big<\Delta \hat{n}_{0}(m)\big> \approx -\frac{160\pi^{2}m^{2}}{9}.
\end{equation}
We note that the number of cycles, $m$, counts the nonlinear phase shift modulus $2\pi$, which is technically 8 times the soliton period~\cite{HH}. The full solution to Eq.~(\ref{timeexp}) takes the form of a quadratic polynomial in $m$, with a negative coefficient of order $10^{-1}$ for the linear term. This linear term may be neglected when $m>10^{-3}$ -- this will be assumed to be the case for the remainder of this work. This perturbative result is valid whenever $|\big< \Delta \hat{n}_{0}(t)\big>| \ll n_{0}.$

The above calculation shows that the change in photon number is a decreasing monotonic function and thus the soliton decays. An important point to note in performing the calculation of Eq.~(\ref{timeexp}) is that the terms with no $\Delta \hat{v}_{c}(x,t)$-dependence are the dominant contributors to the solution and are solely responsible for the coefficient in Eq.~(\ref{Solution}), while the integrals involving $\Delta \hat{v}_{c}(x,t)$ gave negligible contributions and are responsible for the neglected linear term. Moreover, the numerical coefficients which result from the integrals involving $\Delta \hat{v}_{c}(x,t)$ in Eq.~(\ref{timeexp}) are four orders of magnitude smaller than those terms which do not involve $\Delta \hat{v}_{c}(x,t)$. Thus, source terms which involve $\Delta \hat{v}_{c}(x,t)$ may be neglected when one considers soliton cycles which satisfy $m>10^{-3}$ -- as was alluded to in the preceding paragraph. 

Using Eq.~(\ref{photonnumber}) and the relevant expectation values found in Appendix A, one may show that the variance takes on the following form:
\begin{align}\label{VAR}
    \begin{split}
        \sigma^{2}_{\Delta \hat{n}(t_{m})} = n_{0}+58\pi^{2}m^{2}+513\pi^{4}m^{4}.
    \end{split}
\end{align}
The soliton cycle is related to the propagation distance, $z$, along the fiber as~\cite{Haus4}:
\begin{equation}\label{relate}
    m = \frac{1}{4\pi\zeta^{2}_{0}}\frac{k''}{|k'|^{2}}z,
\end{equation}
where $k''$ is the group velocity dispersion (GVD) parameter, $k'$ is the inverse group velocity, and $\zeta_{0}$ is related to the initial soliton width.

As an example, consider a $1$ps pulse with GVD parameter $k''=20 \text{ps}^{2}/\text{km}$ and inverse group velocity $k'=3/(2c)$, where $c$ is the speed of light in vacuum. Combining Eq.~(\ref{Solution}) and Eq.~(\ref{relate}) the soliton must propagate a distance of approximately $47$ meters to lose its first photon.

From the conservation of particle number in Eq.~(\ref{linnumb}), we see that what is lost from Eq.~(\ref{Solution}), must be gained by the terms in Eq.~(\ref{extra}). In the next section, we calculate the gain of the continuum portion of Eq.~(\ref{extra}).

\section{Continuum Radiation Power Spectrum}
Here, as was done in the previous section, we consider the effects of the second order contributions of $\hat{v}$, but now with the purpose of calculating how the generated continuum radiation grows as a function of soliton cycles. We find that the continuum contribution, in comparison to the other three terms in Eq.~(\ref{extra}), makes up only a small portion of the total photons lost, as calculated in Eq.~(\ref{Solution}). More over, it can be shown that the majority of the soliton's photon loss is accounted for by the first term in Eq.~(\ref{extra}), namely:
\begin{equation}
\Delta \hat{n}_{sol} = \frac{2}{n_{0}|c|}\int dx \Big<\Delta\hat{v}^{\dagger}_{sol}\Delta\hat{v}_{sol}\Big>.
\end{equation}
This is expected based on the comments made in the previous section regarding the small contribution made by the continuum portion of Eq.~(\ref{Solution}) when compared to the contribution made by the change in soliton parameter operators. We find that the spectrum of the continuum forms a band having band-width $k\in(-n_{0}|c|,n_{0}|c|)$, which is peaked about the initial soliton momentum.

From Eq.~(\ref{linnumb}), the contribution due solely to the continuum radiation is given by:
\begin{equation}
  \Delta \hat{n}_{cont} = \frac{2}{n_{0}|c|}\int dx \Big<\Delta  \hat{v}^{\dagger}_{c}(x,t)\Delta \hat{v}_{c}(x,t)\Big>.
\end{equation}
Away from the soliton, the continuum radiation field operator, Eq.~(\ref{lintimecon}), takes the form:
\begin{equation}
    \Delta \hat{v}_{c}(x>x_{sw},t) \approx \int_{-\infty}^{\infty}dk e^{-ikx}(k-i)^{2}\hat{a}(k,t),
\end{equation}
where $x_{sw}$ is the characteristic width of the initial soliton pulse.

Thus we may calculate the power spectrum of the emitted continuum radiation by evaluating:
\begin{equation}\label{PowerSpectrum1}
    P(k) \approx \frac{4\pi}{n_{0}|c|}(1+k^{2})^{2}\big<\hat{a}^{\dagger}(k,t)\hat{a}(k,t)\big>.
\end{equation}
Similarly to what we saw in Eq.~(\ref{photonnumber}), including the non-linear term and utilizing the orthogonality relations given in Appendix A, results in the equation of motion for the operator $\hat{a}(k,t)$ to take the following form:
\begin{equation}\label{EOMa}
    \begin{split}
        &\frac{d\hat{a}}{dt} = -i\frac{n_{0}^{2}|c|^{2}}{4}(1+k^{2})\hat{a} + \frac{n_{0}^{2}|c|^{2}}{8}\int dx \bar{\Vec{f}}_{k}^{\dagger}\widehat{NL}\Vec{v},
    \end{split}
\end{equation}
with 
\begin{equation}\label{NL1}
\begin{split}
        &\int dx \bar{\Vec{f}}_{k}^{\dagger}\widehat{NL}\Vec{v} = \\ &\frac{4i}{n_{0}|c|^{1/2}{\pi}}\int dx \frac{\sech(x)e^{ikx}}{(1+k^{2})^{2}}\Big[(2\hat{v}^{\dagger}\hat{v}+\hat{v}^{2})\\&\times(k+i\tanh(x))^2+ \sech^{2}(x)(2\hat{v}^{\dagger}\hat{v} + \hat{v}^{\dagger\,2})\Big].
\end{split}
\end{equation}
Once again, taking an iterative perturbative approach to solve this equation, we insert the solution to the linear problem, Eq.~(\ref{timeevol}), into the right hand side of Eq.~(\ref{EOMa}). To simplify the calculation, we take note of the comments in the previous section where it was pointed out that the terms which involve $\Delta \hat{v}_{c}(x,t)$ in the driving source do not contribute in a meaningful way in comparison to those terms solely involving terms from Eq.~(\ref{vsol}). This applies equally well to Eq.~(\ref{NL1}), hence with this observation, we drop $\Delta \hat{v}_{c}(x,t)$ out of $\hat{v}_{lin}(x,t)$ and use only the modified soliton parameter operators as the driving source.

Equation (\ref{NL1}) evaluates to the following form:
\begin{equation}\label{NL1Solution}
\begin{split}
        &\int dx \bar{\Vec{f}}_{k}^{\dagger}\widehat{NL}\Vec{v} = \hat{f_{1}}(k)+t\hat{f_{2}}(k)+t^{2}\hat{f_{3}}(k),\end{split}
\end{equation}
where $\hat{f_{1}}(k)$, $\hat{f_{2}}(k)$, and $\hat{f_{3}}(k)$ are given in Appendix C as Eqs.~(\ref{f1},\ref{f2},\ref{f3}).

From the form of Eq.~(\ref{NL1Solution}), the solution to Eq.~(\ref{EOMa}) is given as:
\begin{equation}
    \begin{split}
         &\hat{a}(k,t) = \frac{\exp{-i\alpha t}}{\alpha^{3}}\Big[\alpha^{3}\hat{a}(k,0)\\&+\frac{n_{0}^{2}|c|^{2}}{8}\Big(1-\exp{i\alpha t}\Big)\\&\times \Big(i\alpha^{2}\hat{f_{1}}(k)-2i\hat{f_{3}}(k)-\alpha \hat{f_{2}}(k)\Big)\\&+\frac{n_{0}^{2}|c|^{2}}{8}t\exp{i\alpha t} \times\Big(2\alpha \hat{f_{3}}(k)-i\alpha^{2}\hat{f_{2}}(k)\Big)\\&-i\frac{n_{0}^{2}|c|^{2}}{8}\alpha^{2}t^{2}\exp{i\alpha t}\hat{f_{3}}(k)\Big],
    \end{split}
\end{equation}
where
\begin{equation}\label{alpha}
    \alpha = \frac{n_{0}^{2}|c|^{2}(1+k^{2})}{4}.
\end{equation}
Using Eq.~(\ref{PowerSpectrum1}), we plot the power spectrum below in Fig.~(\ref{fig:1}) for multiple values of soliton cycles. The spectrum grows in amplitude as the soliton propagates, while the width of the power spectrum remains constant.
\begin{figure}[hbt!]
    \centering
    \includegraphics[width=8.3cm]{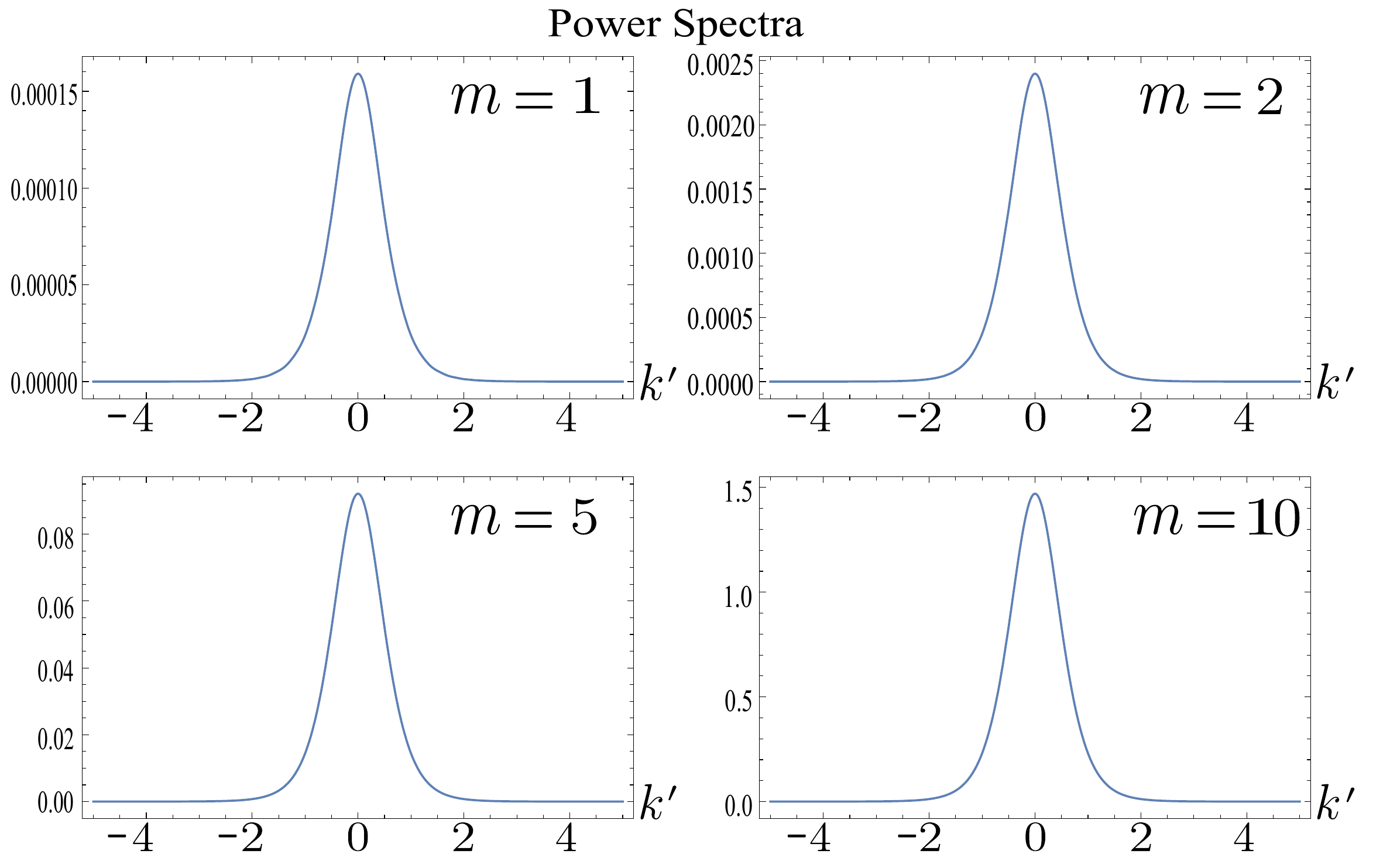}
    \caption{The power spectrum of the continuum radiation for $m=1,2,5,10$ soliton cycles. The spectrum is peaked at the origin and is localized about the band $k'\in(-2,2)$.}
    \label{fig:1}
\end{figure}
The spectrum is peaked at the origin and is localized about the band $k'\in(-2,2)$. Noting the normalized units defined in Eq.~(\ref{eq:relabel}), we see that the spectrum  is peaked about the band:
\begin{equation}
    k\in(-n_{0}|c|,n_{0}|c|).
\end{equation}
The quantity $n_{0}|c|$ appears in Eq.~(\ref{classicalsol}) and is related to the initial width of the soliton pulse. 

\begin{figure}[hbt!]
    \centering
    \includegraphics[width=7.5cm]{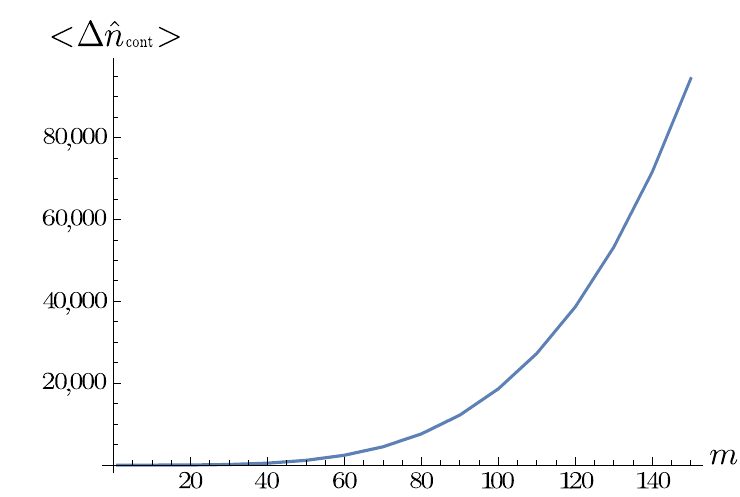}
    \caption{Total continuum contribution with the initial soliton photon number taken to be $n_{0}=10^8$.}
    \label{fig:2}
\end{figure}

Below, in Fig.~(\ref{fig:2}), we plot the total continuum contribution by integrating the power spectrum. One can see that the continuum grows as $m^{4}.$
It is noted that the power spectrum grows as $m^{4},$ while Eq.~(\ref{Solution}) grows as $m^{2}$. This is not an inconsistency -- as was pointed out in the beginning of this section, we have neglected the $\Delta\hat{v}_{sol}$ terms in Eq.~(\ref{extra}). These terms provide the necessary counter terms to account for the total change in photon number, as they must by conservation of Eq.~(\ref{linnumb}). Thus, the amount of photons found solely as continuum radiation is miniscule. For example, comparing Fig.~(\ref{fig:2}) to Eq.~(\ref{Solution}) and considering 100 soliton cycles, one can see that the continuum makes up only $\approx 1\%$ of the total photons lost as calculated in Eq.~(\ref{Solution}). Reiterating what was stated above, the remaining loss of intensity is accounted for in the terms $\Delta\hat{v}^{\dagger}_{sol}\Delta\hat{v}_{sol}$, $\Delta\hat{v}^{\dagger}_{c}\Delta\hat{v}_{sol}$ and $\Delta\hat{v}^{\dagger}_{sol}\Delta\hat{v}_{c}$ as defined in Eq.~(\ref{extra}). 

\section{Conclusion}
By way of performing a perturbation analysis on the linearized quantum non-linear Schr\"{o}dinger equation, we have shown that the optical soliton which is initially in a classical soliton coherent state decays solely due to quantum mechanical effects. It is shown that after $m$ soliton cycles, the soliton photon number decreases proportionally to $m^{2}$, while the generated continuum radiation grows proportional to  $m^{4}$. It is found that the continuum radiation accounts for a small amount of the change of the soliton intensity. The band-width of the continuum radiation lies in the interval $k\in(-n_{0}|c|,n_{0}|c|),$ while being centered about the initial soliton momentum.

Here we make a few comments in regards to how our analysis relates to the work on soliton evaporation presented in Ref.~\cite{Conti}. Our approach has been to expand the quantum soliton field about a classical soliton background, which allows one to analyze the quantum fluctuations of the classical soliton. In contrast, Ref.~\cite{Conti} utilizes the exact solutions of the full quantum field theory of the NLSE to construct a field whose expectation value approximates a classical soliton. The initial state is such that the expectation value of the field results in an average over fundamental soliton solutions having Gaussian distributed momentum and hence having different phase and group velocities, and their dispersion contributes to the evaporation of the soliton.

This work may be expanded on by utilizing the above perturbation analysis to investigate how the quantum evaporation of an optical soliton modifies its squeezing behavior with potential practical implications for precision interferometry~\cite{squeeze1,squeeze2,squeeze3,squeeze4}.

\section{Acknowledgement}

This research is supported by grant number W911NF-19-1-0352 from the United States Army Research Office.

\newpage
\appendix
\section{More Details on the Linearized Theory of the Quantum Soliton} Here we provide a more complete overview of the linearized theory of the quantized NLSE~\cite{Kartner1}. 

The vectors in Eq.~(\ref{expand}) are given as:
\begin{subequations}\label{veq}
\begin{align}
    &\frac{\partial \Vec{\phi}_{cl}(x,t)}{\partial n_{0}}\rvert_{t=0} = \frac{1}{n_{0}}\big[1 - x\tanh{(x)}\big]\phi_{cl}(x,0) \begin{bmatrix} 1 \\ 1 \end{bmatrix},\\&
    \frac{\partial \Vec{\phi}_{cl}(x,t)}{\partial p_{0}}\rvert_{t=0} = \frac{2i}{n_{0}|c|}\,x\,\phi_{cl}(x,0)\begin{bmatrix} 1 \\ -1 \end{bmatrix},\\&
     \frac{\partial \Vec{\phi}_{cl}(x,t)}{\partial x_{0}}\rvert_{t=0} = \big[\frac{n_{0}|c|}{2}\tanh{(x)}\big]\phi_{cl}(x,0)\begin{bmatrix} 1 \\ 1 \end{bmatrix},\\&
      \frac{\partial \Vec{\phi}_{cl}(x,t)}{\partial \theta_{0}}\rvert_{t=0} = i\phi_{cl}(x,0)\begin{bmatrix} 1 \\ -1 \end{bmatrix}.
\end{align}
\end{subequations}
We will abbreviate the above vector equations as $\Vec{\phi}_{i},$ with $i = n_{0},p_{0},x_{0},\theta_{0}.$

The continuum radiation portion of the operator may be expanded as:
\begin{equation}\label{continnum}
    \Delta \Vec{v}_{c}(x,t) = \int^{\infty}_{-\infty}dk\big[\Vec{f}_{k}\,\hat{a}(k,t)+\Vec{g}_{k}\,\hat{b}(k,t)\big],
\end{equation}
where
\begin{equation}\label{con1}
    \Vec{f}_{k} = e^{-ikx}\begin{bmatrix} (k-i\tanh(x))^{2} \\ \sech^{2}(x) \end{bmatrix},
\end{equation}
and
\begin{equation}\label{con2}
    \Vec{g}_{k} = e^{-ikx}\begin{bmatrix}  \sech^{2}(x) \\ (k-i\tanh(x))^{2}  \end{bmatrix}.
\end{equation}

The linear operator defined in Eq.~(\ref{linearop}) acts on the vector equations,  Eqs.~(\ref{veq},\ref{con1},\ref{con2}), as follows:
\begin{subequations}
\begin{align}
&\Vec{L}\Vec{\phi}_{n_{0}} = \frac{2}{n_{0}}\Vec{\phi}_{\theta_{0}},\\&
 \Vec{L}\Vec{\phi}_{p_{0}} = \frac{8}{n_{0}^{2}|c|^{2}}\Vec{\phi}_{x_{0}},\\&
 \Vec{L}\Vec{\phi}_{x_{0}} = 0,\\&
 \Vec{L}\Vec{\phi}_{\theta_{0}} = 0,\\&
 \Vec{L}\Vec{f}_{k} = -i(1+k^{2})\Vec{f}_{k},\\&
 \Vec{L}\Vec{g}_{k} = i(1+k^{2})\Vec{g}_{k}.
 \end{align}
\end{subequations}

We now introduce an inner-product defined as:
\begin{equation}\label{scalar}
   \bra{u}\ket{v} = \frac{1}{2}\int dx u^{\dagger}v.
\end{equation}
With this definition of inner product, one may find a set of functions which give rise to useful orthogonality relations with respect to Eqs.~(\ref{veq},\ref{con1},\ref{con2}). These functions are related through the adjoint of Eq.~(\ref{linearop}). We list them below and denote them with an overbar notation:
\begin{subequations}
\begin{align}
    &\bar{\Vec{\phi}}_{ n_{0}} = \frac{4}{n_{0}|c|}\phi_{cl}(x,0)\begin{bmatrix} 1 \\ 1 \end{bmatrix},\\&
    \bar{\Vec{\phi}}_{ p_{0}} = \frac{2i}{n_{0}}\tanh(x)\phi_{cl}(x,0)\begin{bmatrix} 1 \\ -1 \end{bmatrix},\\&
    \bar{\Vec{\phi}}_{x_{0}} = \frac{8}{n^{3}_{0}|c|^{2}}\,x\,\phi_{cl}(x,0)\begin{bmatrix} 1 \\ 1 \end{bmatrix},\\&
    \bar{\Vec{\phi}}_{\theta_{0}} = \frac{4i}{n^{2}_{0}|c|}(1-x\tanh(x))\phi_{cl}(x,0)\begin{bmatrix} 1 \\ -1 \end{bmatrix},\\&
    \bar{\Vec{f}}_{k} = \frac{e^{-ikx}}{\pi(1+k^{2})^{2}}\begin{bmatrix} (k-i\tanh(x))^{2} \\ -\sech^{2}(x) \end{bmatrix},\\&
    \bar{\Vec{g}}_{k} = \frac{e^{-ikx}}{\pi(1+k^{2})^{2}}\begin{bmatrix} \sech^{2}(x) \\ -(k-i\tanh(x))^{2}  \end{bmatrix}.
\end{align}
\end{subequations}
With the above defined vectors and inner-product, one can show that the following orthogonality relationships hold:
\begin{subequations}
\begin{align}
    &\bra{\bar{\Vec{\phi}}_{i}}\ket{\Vec{\phi}_{j}} = \delta_{ij},\\&
    \bra{\bar{\Vec{\phi}}_{i}}\ket{\Vec{f}_{k}} = \bra{\bar{\Vec{\phi}}_{i}}\ket{\Vec{g}_{k}} = 0,\\&
    \bra{\bar{\Vec{f}}_{k}}\ket{\Vec{f}_{k'}} = \bra{\bar{\Vec{g}}_{k}}\ket{\Vec{g}_{k'}} = \delta(k-k').
\end{align}
\end{subequations}
We may now solve for the time evolution of the perturbation operators. Inserting Eq.~(\ref{expand}) into Eq.~(\ref{eigen}) and multiplying both sides by the proper adjoint equation, while making use of the orthogonality relations through the  inner-product Eq.~(\ref{scalar}), allows for the time evolution of each operator in Eq.~(\ref{expand}) to be isolated.

We arrive at the following equations of motion:
\begin{subequations}
\label{evolpara}
    \begin{align}
        &\frac{4}{n^{2}_{0}|c|^{2}}\frac{d}{dt}\Delta \hat{n}_{0}(t) = 0,\\
        &\frac{4}{n^{2}_{0}|c|^{2}}\frac{d}{dt}\Delta \hat{\theta}_{0}(t) = \frac{2}{n_{0}}\Delta \hat{n}_{0}(0) ,\\
        &\frac{4}{n^{2}_{0}|c|^{2}}\frac{d}{dt}\Delta \hat{p}_{0}(t) = 0,\\
        &\frac{4}{n^{2}_{0}|c|^{2}}\frac{d}{dt}\Delta \hat{x}_{0}(t) = \frac{8}{n^{2}_{0}|c|^{2}}\Delta \hat{p}_{0}(0),
       \\& \frac{4}{n^{2}_{0}|c|^{2}}\frac{d\hat{a}}{dt} = -i(1+k^{2})\hat{a},\\
        &\frac{4}{n^{2}_{0}|c|^{2}}\frac{d\hat{b}}{dt} = i(1+k^{2})\hat{b}.
    \end{align}
\end{subequations}
Thus, from the above equations, the change in photon number and the change in momentum evolve independently of time. The center of soliton position and the phase evolve according to:
\begin{subequations}
\label{lintime}
    \begin{align}
        &\Delta \hat{\theta}_{0}(t) = \Delta \hat{\theta}_{0}(0) + \frac{n_{0}|c|^{2}t}{2}\Delta \hat{n}_{0}(0) ,\\
        &\Delta \hat{x}_{0}(t) = \Delta \hat{x}_{0}(0) + 2\Delta \hat{p}_{0}(0)t,
    \end{align}
\end{subequations}
while the continuum operators evolve as
\begin{equation}\label{contimeevolve}
\begin{split}
    &\hat{a}(k,t) = \hat{a}(k,0)\exp[-i\frac{n_{0}^{2}|c|^{2}}{4}(1+k^{2})t],\\
   &\hat{b}(k,t) = \hat{b}(k,0)\exp[i\frac{n_{0}^{2}|c|^{2}}{4}(1+k^{2})t].
   \end{split}
\end{equation}

\section{Expectation Values}
In this appendix, we list the relevant expectation values that are used in the calculation of Eq.~(\ref{timeexp}) and Eq.~(\ref{PowerSpectrum1}). It is noted that when calculating the expectation value of the product of four operators, the expectation value decomposes into a linear combination of all permutations of products of two-point expectation values -- thus only the expectation values provided below are required for the evaluation of Eq.~(\ref{Solution}), Eq.~(\ref{VAR}) and Eq.~(\ref{PowerSpectrum1}). 

From the linearized theory of the quantum soliton, presented in Section II, one may use the inner-product defined in Eq.~(\ref{scalar}), along with the expansion given in Eq.~(\ref{expand}), to isolate the perturbed soliton parameter operators and the continuum operator. They are given below as:
\begin{align}\label{a1}
    \begin{split}
            &\Delta \hat{n}_{0}(0) = \frac{1}{|c|^{1/2}}\int_{-\infty}^{\infty}dx \sech{(x)}\\&\times\big(\hat{v}(x,0)+\hat{v}^{\dagger}(x,0)\big),
    \end{split}\\
    \begin{split}
      &\Delta \hat{\theta}_{0}(0) = \frac{-i}{n_{0}|c|^{1/2}}\int_{-\infty}^{\infty}dx (1-x\tanh{(x)})\\&\sech{(x)}\times\big(\hat{v}(x,0)-\hat{v}^{\dagger}(x,0)\big),
    \end{split}\\
    \begin{split}
         &\Delta \hat{p}_{0}(0) = \frac{-i|c|^{1/2}}{2}\int_{-\infty}^{\infty}dx \tanh{(x)}\sech{(x)}\times\\&\big(\hat{v}(x,0)-\hat{v}^{\dagger}(x,0)\big),
    \end{split}\\
    \begin{split}
        &\Delta \hat{x}_{0}(0) = \frac{2}{n_{0}^{2}|c|^{3/2}}\int_{-\infty}^{\infty}dx x\sech{(x)}\\&\times\big(\hat{v}(x,0)+\hat{v}^{\dagger}(x,0)\big),
    \end{split}
\end{align}
    and
\begin{align}
    \begin{split}
         &\hat{a}(k,0) = \frac{1}{2}\int_{-\infty}^{\infty}dx \frac{\exp{{ikx}}}{\pi(1+k^{2})^{2}}\times\\&\big[(k+i\tanh{(x)})^{2}\hat{v}(x,0) - \sech{(x)}^{2}\hat{v}^{\dagger}(x,0)\big],
    \end{split}\\
    \begin{split}
        &\hat{b}(k,0) = \frac{1}{2}\int_{-\infty}^{\infty}dx \frac{\exp{{ikx}}}{\pi(1+k^{2})^{2}}\times\\&\big[\sech{(x)}^{2}\hat{v}(x,0)-(k+i\tanh{(x)})^{2}\hat{v}^{\dagger}(x,0)\big].
    \end{split}
\end{align}
From the above six operator equations and making use of the commutation relation $[\hat{v}(x,t),\hat{v}^{\dagger}(y,t)] = \frac{n_{0}|c|}{2}\delta(x-y)$ (the scaling factor in front of the delta function is due to the rescaling introduced in Eq.~(\ref{eq:relabel})), one may calculate the following expectation values on the vacuum state of the linearized theory:
\begin{align}
\begin{split}
&\Big<\Delta \hat{n}_{0}^{2}(0)\Big> = n_{0},
\end{split}\\
\begin{split}
       \Big<\Delta\hat{\theta}_{0}^{2}(0)\Big> = \frac{0.6075}{n_{0}},
    \end{split}\\
    \begin{split}
\Big<\Delta \hat{p}_{0}^{2}(0)\Big> =
\frac{n_{0}|c|^{2}}{12},
    \end{split}\\
    \begin{split}
        \Big<\Delta \hat{x}_{0}^{2}(0)\Big> = \frac{\pi^{2}}{3n_{0}^{3}|c|^{2}},
    \end{split}
\end{align}
\begin{align}
\begin{split}
&\frac{\Big<\hat{a}(k,0)\hat{a}(k',0)\Big>}{n_{0}|c|} =\\&\frac{\left(k'+k\right) \left(2-6 k^2+6 k \left(k'+k\right)-\left(k'+k\right)^2\right)}{48 \pi  \left(k^2+1\right)^2 \left(k'^2+1\right)^2}\\&\times\text{csch}\left(\frac{1}{2} \pi  \left(k'+k\right)\right),
\end{split}
\end{align}
\begin{align}
\begin{split}
&\frac{\Big<\hat{a}(k,0)\hat{b}(k',0)\Big>}{n_{0}|c|} =\\&\delta\left(k+k'\right)\Big(\frac{-k^{2}k'^{2}}{4\pi (1+k^2)^{2}(1+k'^{2})^{2}}\\&+2\pi \big(1-k^{2}-k'^{2}-4kk'\big)\Big)\\&+\text{csch}\left(\frac{1}{2} \pi  \left(k'+k\right)\right)\\&\times\Big(\frac{-kk'(k+k')}{4\pi (1+k^2)^{2}(1+k'^{2})^{2}}+\pi(k+k')(-\frac{10}{3}\\&+\frac{1}{6}(13k^{2}+13k'^{2}+38kk'))\Big),
\end{split}
\end{align}
\begin{align}
    \begin{split}
    &\frac{\Big<\hat{b}(k,0)\hat{a}(k',0)\Big>}{n_{0}|c|} =\\& -\frac{\left(k'+k\right) \left(\left(k'+k\right)^2+4\right) \text{csch}\left(\frac{1}{2} \pi  \left(k'+k\right)\right)}{48 \pi  \left(k^2+1\right)^2 \left(k'^2+1\right)^2},
    \end{split}
\end{align}
\begin{align}
        \begin{split}
        &\frac{\Big<\hat{b}(k,0)\hat{b}(k',0)\Big>}{n_{0}|c|} =\\& -\frac{\left(k'+k\right) \left(6 k^2-6 k \left(k'+k\right)+\left(k'+k\right)^2-2\right)}{48 \pi  \left(k^2+1\right)^2 \left(k'^2+1\right)^2}\\&\times \text{csch}\left(\frac{1}{2} \pi  \left(k'+k\right)\right),
    \end{split}
\end{align}
\begin{align}
    \begin{split}
        \Big<\Delta \hat{n}_{0}(0) \Delta \hat{\theta}_{0}(0)\Big> = -\Big<\Delta \hat{\theta}_{0}(0) \Delta \hat{n}_{0}(0) \Big> = \frac{i}{2},
    \end{split}\\
    \begin{split}
        \Big<\Delta \hat{x}_{0}(0) \Delta \hat{p}_{0}(0)\Big> = -\Big<\Delta \hat{p}_{0}(0) \Delta \hat{x}_{0}(0) \Big> = \frac{i}{2n_{0}},
         \end{split}
\end{align}
\begin{align}
\begin{split}
&\Big<\Delta \hat{n}_{0}(0)\hat{a}(k,0)\Big>=\Big<\hat{a}(k,0)\Delta \hat{n}_{0}(0)\Big> = \\&
-\Big<\Delta \hat{n}_{0}(0)\hat{b}(k,0)\Big>= -\Big<\hat{b}(k,0)\Delta \hat{n}_{0}(0)\Big>\\& =-\frac{n_{0}|c|^{1/2}}{8}\Big[\frac{\sech{(\frac{k\pi}{2})}}{(1+k^{2})}\Big],
\end{split}
\end{align}
\begin{align}
\begin{split}
        &\Big<\Delta \hat{\theta}_{0}(0)\hat{a}(k,0)\Big>=\Big<\hat{a}(k,0)\Delta \hat{\theta}_{0}(0)\Big> =\\&\Big<\Delta \hat{\theta}_{0}(0)\hat{b}(k,0)\Big>= \Big<\hat{b}(k,0)\Delta \hat{\theta}_{0}(0)\Big> = \\&\frac{i|c|^{1/2}}{48}\Big[\frac{\sech{(\frac{k\pi}{2})}\Big(4+(k+k^{3})\pi\tanh{(\frac{k\pi}{2})}\Big)}{(1+k^{2})^{2}}\Big],
\end{split}
\end{align}
\begin{align}
\begin{split}
    &\Big<\Delta \hat{n}_{0}(0)\hat{a}(k,0)\Big>=\Big<\hat{a}(k,0)\Delta \hat{n}_{0}(0)\Big> = \\&
    -\Big<\Delta \hat{n}_{0}(0)\hat{b}(k,0)\Big>= -\Big<\hat{b}(k,0)\Delta \hat{n}_{0}(0)\Big>\\& =-\frac{n_{0}|c|^{1/2}}{8}\Big[\frac{\sech{(\frac{k\pi}{2})}}{(1+k^{2})}\Big],
\end{split}
\end{align}
\begin{align}
\begin{split}
    &\Big<\Delta \hat{\theta}_{0}(0)\hat{a}(k,0)\Big>=\Big<\hat{a}(k,0)\Delta \hat{\theta}_{0}(0)\Big> =\\&\Big<\Delta \hat{\theta}_{0}(0)\hat{b}(k,0)\Big>= \Big<\hat{b}(k,0)\Delta \hat{\theta}_{0}(0)\Big> = \\&\frac{i|c|^{1/2}}{48}\Big[\frac{\sech{(\frac{k\pi}{2})}\Big(4+(k+k^{3})\pi\tanh{(\frac{k\pi}{2})}\Big)}{(1+k^{2})^{2}}\Big],
\end{split}
\end{align}
\begin{align}
\begin{split}
    &\Big<\Delta \hat{p}_{0}(0)\hat{a}(k,0)\Big>=\Big<\hat{a}(k,0)\Delta \hat{p}_{0}(0)\Big> = \\&
    \Big<\Delta \hat{p}_{0}(0)\hat{b}(k,0)\Big>= \Big<\hat{b}(k,0)\Delta \hat{p}_{0}(0)\Big> = \\&
    -\frac{n_{0}|c|^{3/2}}{48}\Big(\frac{k\sech{(\frac{k\pi}{2})}}{(1+k^{2})}\Big),
\end{split}
\end{align}
\begin{equation}
\begin{split}
    &\Big<\Delta \hat{x}_{0}(0)\hat{a}(k,0)\Big>=\Big<\hat{a}(k,0)\Delta \hat{x}_{0}(0)\Big> =
    \\&-\Big<\Delta \hat{x}_{0}(0)\hat{b}(k,0)\Big>= -\Big<\hat{b}(k,0)\Delta \hat{x}_{0}(0)\Big> =
    \\&\frac{-i}{8n_{0}|c|^{1/2}}\Big(\frac{\big(-4k+(1+k^{2})\pi\tanh{(\frac{k\pi}{2})}\big)}{(1+k^{2})^{2}}\Big)\\&\times\sech{(\frac{k\pi}{2})},
    \end{split}
\end{equation}
and 
\begin{align}
\begin{split}
        &\Big<\Delta \hat{n}_{0}(0)\Delta \hat{p}_{0}(0)\Big>= \Big<\Delta \hat{p}_{0}(0)\Delta \hat{n}_{0}(0)\Big> = 0,
\end{split}\\
\begin{split}
        &\Big<\Delta \hat{n}_{0}(0)\Delta \hat{x}_{0}(0)\Big> = \Big<\Delta \hat{x}_{0}(0)\Delta \hat{n}_{0}(0)\Big> = 0,
\end{split}\\
\begin{split}
         &\Big<\Delta \hat{p}_{0}(0)\Delta \hat{\theta}_{0}(0)\Big> = \Big<\Delta \hat{\theta}_{0}(0)\Delta \hat{p}_{0}(0)\Big> = 0,
\end{split}\\
\begin{split}
        &\Big<\Delta \hat{x}_{0}(0)\Delta \hat{\theta}_{0}(0)\Big> = \Big<\Delta\hat{\theta}_{0}(0)\Delta \hat{x}_{0}(0)\Big> =0.
\end{split}
\end{align}
\section{Functions for Continuum Radiation Power Spectrum}

Here we explicitly give the functions $\hat{f_{1}}(k)$, $\hat{f_{2}}(k)$, and $\hat{f_{3}}(k)$ that were introduced in Eq.~(\ref{NL1Solution}). They are given as:
\begin{equation}\label{f1}
\hat{f}_{1}(k) = \hat{g}_{1}(k)+\hat{h}_{1}(k)+\hat{j_{1}}(k),
\end{equation}
where
\begin{equation}\label{g1}
\begin{split}
 &\hat{g}_{1}(k)= \\&  \frac{4i}{n_{0}|c|^{1/2}\pi}\Big[ \frac{3|c|\Delta \hat{n}^{2}(0)}{1152 \left(k^2+1\right)^2}\Big(\pi\sech^{3}{(\frac{k\pi}{2})}\\&\Big(32-3k^{2}\pi^{2}-6k^{4}\pi^{2}-3k^{6}\pi^{2}\\&+(32+k^{2}\pi^{2}+2k^{4}\pi^{2}+k^{6}\pi^{2})\cosh{(k\pi)}\\&+16k(1+k^{2})\pi\sinh{(k\pi)}\Big) \Big)\\&+\frac{n_{0}^{2}|c|\Delta \hat{\theta}^{2}(0)}{16}\pi\sech{(\frac{k\pi}{2})}-\\&\frac{\Delta \hat{p}^{2}(0)}{96|c|(1+k^{2})^{2}}\pi\sech^{3}{(\frac{k\pi}{2})}\Big(32-9\pi^{2}-18k^{2}\pi^{2}\\&-9k^{4}\pi^{2}+(32+3(1+k^{2})^{2}\pi^{2})\cosh{(k\pi)}\\&-16k(1+k^{2})\pi\sinh{(k\pi)}\Big)\\&-\frac{3n_{0}^{4}|c|^{3}\Delta \hat{x}^{2}(0)}{576}k^2\pi \sech{(\frac{k\pi}{2})}\\&+\frac{n_{0}|c|\Delta \hat{n}(0)\Delta\hat{\theta}(0)}{480(1+k^{2})^{2}}\times\\&i\pi\sech{(\frac{k\pi}{2})}(24-40k^{2}+k(33+50k^{2}+17k^{4})\\&\times\pi\tanh{({\frac{\pi k}{2}})})\\&-\frac{n_{0}|c|\Delta\hat{\theta}(0)\Delta \hat{n}(0)}{480(1+k^{2})^{2}} i\pi\sech{(\frac{k\pi}{2})} (8(13+5k^{2})\\&+k(23+30k^{2}+7k^{4})\pi\tanh{({\frac{\pi k}{2}})})+\\& \frac{n_{0}^{2}|c|^{2}\acomm{\Delta\hat{n}(0)}{\Delta\hat{x}(0)}}{192(1+k^{2})}i\pi k\sech{(\frac{k\pi}{2})}(8+\\&k(1+k^{2})\pi\tanh{(\frac{k\pi}{2})})\Big],
\end{split}
\end{equation}
\begin{equation}\label{h1}
\begin{split}
 &\hat{h}_{1}(k)= \\& \frac{4i}{n_{0}|c|^{1/2}\pi}\Big[\frac{\Delta \hat{n}(0)\Delta\hat{p}(0)}{960(1+k^2)^2}\Big(\pi\sech^{3}{(\frac{k\pi}{2})}\Big(k(-160\\&+3(33+50k^{2}+17k^{4})\pi^{2})
 \\&-k(160+(33+50k^{2}+17k^{4})\pi^{2})\cosh{(k\pi)}\\&+16(-3+10k^{2}+5k^{4})\pi\sinh{(k\pi)}\Big) - \\&\frac{\Delta \hat{p}(0)\Delta\hat{n}(0)}{960(1+k^2)^2}\Big(\pi\sech^{3}{(\frac{k\pi}{2})}\Big(k(160+\\&3(23+30k^{2}+7k^{4})\pi^{2})\\&-k(-160+(23+30k^{2}+7k^{4})\pi^{2})\cosh{(k\pi)}\\&+8(-11+10k^{2}+5k^{40})\pi\sinh{(k\pi)}\Big)\\&-\frac{n_{0}^{3}|c|^{2}\Delta \hat{x}(0)\Delta\hat{\theta}(0)}{480(1+k^{2})}k(33+17k^{2})\pi\sech{(\frac{k\pi}{2})}\Big]
\end{split}
\end{equation}
and
\begin{equation}\label{j1}
\begin{split}
  &\hat{j_{1}}(k) = \\&  \frac{4i}{n_{0}|c|^{1/2}\pi}\Big[\frac{n_{0}^{3}|c|^{2}\Delta \hat{\theta}(0)\Delta\hat{x}(0)}{480(1+k^{2})}k(23+7k^{2})\\&\times\pi\sech{(\frac{k\pi}{2})}+\frac{n_{0}\acomm{\Delta\hat{\theta}(0)}{\Delta\hat{p}(0)}}{48(1+k^{2})}\\&\times i\pi \sech{(\frac{k\pi}{2})}(-8k+3(1+k^{2})\pi\tanh{(\frac{k\pi}{2})})\\&-\frac{n_{0}^{2}|c|\Delta \hat{x}(0)\Delta\hat{p}(0)}{480(1+k^{2})}i\pi \sech{(\frac{k\pi}{2})}(-8(-3+\\&15k^{2}+10k^{4})+k(33+50k^{2}+17k^{4})\pi\tanh{(\frac{k\pi}{2})})\\&+\frac{n_{0}^{2}|c|\Delta \hat{p}(0)\Delta\hat{x}(0)}{480(1+k^{2})}\times\\&i\pi \sech{(\frac{k\pi}{2})}(-8(2+15k^{2}+5k^{4})\\&+k(23+30k^{2}+7k^{4})\pi\tanh{(\frac{k\pi}{2})})\Big].
  \end{split}
\end{equation}
\begin{equation}\label{f2}
    \begin{split}
    &\hat{f_{2}}(k) = \frac{4i}{n_{0}|c|^{1/2}{\pi}}\Big[\frac{n_{0}^{2}|c|^{3}\Delta \hat{n}^{2}(0)}{96(1+k^{2})}i\pi \sech{(\frac{k\pi}{2})}(-8\\&+k(1+k^{2})\pi\tanh{(\frac{k\pi}{2})})\\&-\frac{n_{0}^{2}|c|\Delta \hat{p}^{2}(0)}{24(1+k^{2})}i\pi \sech{(\frac{k\pi}{2})}(4-4k^{2}\\&+k(1+k^{2})\pi\tanh{(\frac{k\pi}{2})})\\&+\frac{n_{0}^{2}|c|^{2}\acomm{\Delta\hat{n}(0)}{\Delta\hat{p}(0)}}{1920(1+k^{2})^{2}}\times\\&i\pi \sech{(\frac{k\pi}{2})}(-512k+(135+161k^{2}+45k^{4}\\&+19k^{6})\pi\tanh{(\frac{k\pi}{2})})\\&+\frac{n_{0}^{3}|c|^{3}\acomm{\Delta\hat{n}(0)}{\Delta\hat{\theta}(0)}}{32}\pi\sech{(\frac{k\pi}{2})}\\&-\frac{n_{0}^{3}|c|^{2}\Delta \hat{p}(0)\Delta\hat{\theta}(0)}{240(1+k^{2})}\times\pi\sech{(\frac{k\pi}{2})}k(33+17k^{2})\\&+\frac{n_{0}^{3}|c|^{2}\Delta \hat{\theta}(0)\Delta\hat{p}(0)}{240(1+k^{2})}\pi\sech{(\frac{k\pi}{2})}k(23+7k^{2})\\&-\frac{n_{0}^{4}|c|^{4}\Delta \hat{x}(0)\Delta\hat{n}(0)}{960(1+k^{2})}\pi\sech{(\frac{k\pi}{2})}k(33+17k^{2})\\&+\frac{n_{0}^{4}|c|^{4}\Delta \hat{n}(0)\Delta\hat{x}(0)}{960(1+k^{2})}\times\pi\sech{(\frac{k\pi}{2})}k(23+7k^{2})\\&-\frac{n_{0}^{4}|c|^{3}\acomm{\Delta\hat{p}(0)}{\Delta\hat{x}(0)}}{96}\pi k^{2}\sech{(\frac{k\pi}{2})}\Big]
    \end{split}
\end{equation}
and 
\begin{equation}\label{f3}
    \begin{split}
    &\hat{f_{3}}(k) = \frac{4i}{n_{0}|c|^{1/2}{\pi}}\Big[\frac{n_{0}^{4}|c|^{5}\Delta\hat{n}^{2}(0)}{64}\pi\sech{(\frac{k\pi}{2})}\\&-\frac{n_{0}^{4}|c|^{3}\Delta\hat{p}^{2}(0)}{48}\pi k^{2}\sech{(\frac{k\pi}{2})}-\\&\frac{n_{0}^{4}|c|^{4}\Delta \hat{p}(0)\Delta\hat{n}(0)}{480(1+k^{2})}\times k(33+17k^{2})\pi\sech{(\frac{k\pi}{2})}\\&+\frac{n_{0}^{4}|c|^{4}\Delta \hat{n}(0)\Delta\hat{p}(0)}{480(1+k^{2})}\times k(23+7k^{2})\pi\sech{(\frac{k\pi}{2})}\Big].
    \end{split}
\end{equation}

\end{document}